\documentclass[nofootinbib,prd,twocolumn,showpacs,showkeys,preprintnumbers]{revtex4-1}
\usepackage{hyperref,amssymb,amsmath,mathrsfs,bm,graphicx}
\usepackage{enumitem}
\input{epsf}
\begin{document}
\title {Geodesics of the hyperbolically symmetric black hole}
\author{L. Herrera}
\email{lherrera@usal.es}
\affiliation{Instituto Universitario de F\'isica
Fundamental y Matem\'aticas, Universidad de Salamanca, Salamanca 37007, Spain}
\author{A. Di Prisco}
\email{alicia.diprisco@ciens.ucv.ve}
\affiliation{Escuela de F\'\i sica, Facultad de Ciencias, Universidad Central de Venezuela, Caracas 1050, Venezuela}
\author{J. Ospino}
\email{j.ospino@usal.es}
\affiliation{Departamento de Matem\'atica Aplicada and Instituto Universitario de F\'isica
Fundamental y Matem\'aticas,, Universidad de Salamanca, Salamanca 37007, Spain}
\author{Louis Witten}
\email{lwittenw@gmail.com}
\affiliation{Department of Physics, University of Florida, Gainesville, FL. 32611, USA}

\date{\today}
\begin{abstract}
We carry out a systematic study on the motion of test particles in the region inner to the horizon of a hyperbolically symmetric black hole. The geodesic equations are written and analyzed in detail. The obtained results are contrasted with the corresponding results obtained for the spherically symmetric case. It is found that test particles experience a repulsive force within the horizon, which prevents them to reach the center. These results are obtained for radially moving particles as well as for particles moving in the $\theta-R$ subspace. To complement our study we calculate the precession of a gyroscope moving along  a circular path (non--geodesic) within the horizon. We obtain that  the precession of the gyroscope is retrograde in the rotating frame, unlike the precession close to the horizon ($R=2m+\epsilon$) in the Schwarzschild spacetime, which is forward.
\end{abstract}
\date{\today}
\maketitle
\section{Introduction}
In a recent paper \cite{1} a global description of the  Schwarzschild black hole was proposed, which  sharply differs from the ``classical''  picture of the spherically symmetric black hole. The motivation for this proposal was based on the well known fact that any transformation that maintains the static form of the Schwarzschild metric (in the whole space--time) is unable to remove the singularity in the line element  \cite{rosen}. In other words, 
any  coordinate transformation  allowing the manifold to extend over the whole space--time (including the region inner to the horizon),  necessarily  implies that the metric is  non-static within the horizon (see for example \cite{eddington, 1b, fin, krus, is}). A simple way to arrive at  this  conclusion consists in noticing that the Schwarzschild horizon is also a Killing horizon, implying that the time--like Killing vector outside the horizon becomes space--like inside it. If we recall that a static observer is one whose four--velocity  is proportional to the Killing time--like vector \cite{Caroll}, it follows that no static observers can be defined inside the horizon. Further discussion on this point may be found  in \cite{Rin}.

Then, based on the physical point of view that  any equilibrium final state of a physical process should  be static, the existence of a static solution would be expected over the whole space--time. To  achieve that, the following scheme was proposed in \cite{1}.

Outside the horizon ($R >2 m$), one has the usual Schwarzschild  line element corresponding to the spherically symmetric vacuum solution to Einstein  equations, which  can be written in polar coordinates in the form 
\begin{eqnarray}
ds^2&=&-\left(1-\frac{2m}{R}\right)dt^2+\frac{dR^2}{\left(1-\frac{2m}{R}\right)}+R^2d\Omega^2, \nonumber \\ d\Omega^2&=&d\theta^2+\sin^2 \theta d\phi^2.
\label{w2}
\end{eqnarray}

As is well known, this metric is static and spherically symmetric, meaning that it admits four Killing vectors:
\begin{eqnarray}
\mathbf{\chi }_{(\mathbf{0})} = \partial _{\mathbf{t}}, \quad {\bf \chi_{(2)}}=-\cos \phi \partial_{\theta}+\cot\theta \sin\phi \partial_{\phi}\nonumber \\
{\bf \chi_{(1)}}=\partial_{\phi} \quad {\bf \chi_{(3)}}=\sin \phi \partial_{\theta}+\cot\theta \cos\phi \partial_{\phi}.
\label{2cmh}
\end{eqnarray}

However, when $R <2 m$ the signature changes from (-, +, +, +) to (+, -, +, +) and an apparent line element singularity  appears at  $R =2 m$. Of course, as is also well known,  these drawbacks  can be removed by coordinate transformations, but at the price that, as mentioned before, the staticity is lost within the horizon.

In order to save the staticity inside  the horizon, the   model proposed in \cite{1} describes the space time as consisting of a complete four dimensional manifold  (described by (\ref{w2})) on the exterior side and a second complete four dimensional solution in the interior.  Additionally  a change in signature, as well as a  change in the symmetry at the horizon was required. The $\theta-\phi$ sub-manifolds have a spherical symmetry on the exterior and hyperbolic symmetry in the interior. The two meet only at  $R=2m$, $\theta=0$. 

Thus the model  permits for a change in symmetry, from spherical outside the horizon to hyperbolic inside the horizon. Doing so, one has a static solution everywhere, but the symmetry of the
$R =2 m$ surface is different at both sides of it. We have to stress that we  do not know if there is any specific mechanism behind such a change of symmetry and signature. However,  the main point is that  the change of symmetry (and signature) was the only way we have found to obtain a globally static solution.

Thus, the solution proposed for $R <2 m$ is:
\begin{eqnarray}
ds^2&=&\left(\frac{2m}{R}-1\right)dt^2-\frac{dR^2}{\left(\frac{2m}{R}-1\right)}-R^2d\Omega^2, \nonumber \\ d\Omega^2&=&d\theta^2+\sinh^2 \theta d\phi^2.
\label{w3}
\end{eqnarray}

This is a static solution with the $(\theta  ,\phi )$ space describing a positive Gaussian curvature. 

Besides the time--like Killing vector $\mathbf{\chi }_{(\mathbf{0})} = \partial _{\mathbf{t}}$, it admits three additional Killing vectors which are:
\begin{eqnarray}
{\bf \chi_{(1)}}&=&\partial_{\phi},\qquad  {\bf \chi_{(2)}}=-\cos \phi \partial_{\theta}+\coth\theta \sin\phi \partial_{\phi},\nonumber \\ {\bf \chi_{(3)}}&=&\sin \phi \partial_{\theta}+\coth\theta \cos\phi \partial_{\phi}.
\label{2cmhb}
\end{eqnarray}

A solution to the Einstein equations of the form given by (\ref{w3}), defined by the hyperbolic symmetry  (\ref{2cmhb}), was first considered by Harrison \cite{Ha}, and has been more recently the subject of research  in different contexts (see \cite{ellis, 1n, Ga, Ri, Ka, Ma, ren} and references therein).

In \cite{1}, a general study of radial geodesic at $\theta=0$ was presented, leading to some interesting conclusions about the behaviour of  a test   particle in this new picture of the Schwarzschild  black hole. Our purpose in this work is to carry out  a complete study on the geodesics in the region inner to the horizon. Furthermore, some erroneous conclusions  about the motion of test particles along the $\theta=0$ axis presented in \cite{1}, will be corrected.

As we shall see here, very important differences appear in the behaviour of test particles inside the horizon, when this region is described by (\ref{w3}), as compared with the results obtained for  the ``classical'' black hole picture. Particularly relevant are the facts that a repulsive acceleration is experienced by the test particle inside the horizon, and that test particles can cross the horizon outwardly, but only along the axis $\theta=0$.

\section{The Geodesics}
The equations governing the geodesics may be derived from the Lagrangian

\begin{equation}
2{\cal L}=g_{\alpha\beta}\dot{x}^\alpha\dot{x}^\beta,
\label{l1}
\end{equation}
where the dot denotes differentiation with respect to an affine parameter
$s$, which for timelike geodesics coincides with the proper time. Then,
 the Euler-Lagrange equations,
\begin{equation}
\frac{d}{ds}\left(\frac{\partial{\cal
L}}{\partial\dot{x}^\alpha}\right)-\frac{\partial{\cal L}}
{\partial x^\alpha}=0,
\label{l2}
\end{equation}
lead to the geodesic equations, which may also be written  in its usual form,

\begin{equation}
\ddot x^\alpha+\Gamma^\alpha_{\beta \gamma} \dot x^\beta \dot x^\gamma=0.
\label{g11}
\end{equation}

Although the general characteristics   of geodesics for $R>2m$ are very well known, here we include a very brief resume, in order  to contrast these with the results that we shall obtain for $R<2m$.
\subsection{Geodesics for $R>2m$ (Schwarzschild)}
For the metric (\ref{w2}) the geodesic equations (\ref{l2}) are,

\begin{equation}
\ddot t+\frac{2m\dot t \dot R}{R^2(1-\frac{2m}{R})}=0
\label{gs1}
\end{equation}

\begin{equation}
\frac{\ddot R}{1-\frac{2m}{R}}-\frac{\dot R^2 m}{R^2(1-\frac{2m}{R})^2}+\frac{m \dot t^2}{R^2}-R \dot \theta^2-R\dot \phi ^2 \sin^2 \theta=0,
\label{gs2}
\end{equation}

\begin{equation}
\ddot \theta R^2+2R\dot R \dot \theta-R^2\dot \phi ^2  \sin\theta \cos \theta=0,
\label{gs3}
\end{equation}

\begin{equation}
\ddot \phi \sin^2\theta +\frac{2\dot R \dot \phi\sin^2\theta}{R}+2\dot \phi \dot \theta \sin\theta \cos \theta=0.
\label{gs4}
\end{equation}
\\
As is well known, there are unbounded orbits as well as bounded ones. In this latter case we have elliptic orbits with a perihelion shift. There are also circular orbits, which may be stable or unstable. 

Let us first consider circular geodesics ($\dot R=\dot \theta=0$), then it follows from (\ref{gs1}) and (\ref{gs4}) that $\ddot t=\ddot \phi=0$, and from (\ref{gs3}) we obtain
\begin{equation}
R^2\dot \phi ^2  \sin\theta \cos \theta=0,
\label{gs3b}
\end{equation}
which implies that circular geodesics do exist  on the plane $\theta=\pi/2$. Of course, due to the spherical symmetry, if the particle is not on this plane we can always rotate coordinates until it is. Accordingly without loss of generality we may choose $\theta=\pi/2$.

Next, from  (\ref{gs2}) we obtain

\begin{equation}
\frac{m \dot t^2}{R^2}-R\dot \phi ^2 =0,
\label{gs2b}
\end{equation}
then defining the angular velocity as $\omega=\frac{\dot \phi}{\dot t}$ we obtain the Kepler law
\begin{equation}
\omega^2=\frac{m}{R^3}.
\label{gs2bc}
\end{equation}

Let us now define a ``velocity'' by \cite{An}:
\begin{equation}
W^i=\frac{dx^i}{\sqrt {-g_{00}}dx^0},
\label{vel}
\end{equation}
with
\begin{equation}
dx^i=(0, dx^1, dx^2, dx^3).
\label{vel1}
\end{equation}

Then for the tangential velocity of a circular orbit we find

\begin{equation}
W^2\equiv \vert W^i  W_i\vert=\omega^2 R^2\left(1-\frac{2m}{R}\right)^{-1}.
\label{vel2}
\end{equation}

In the weak field limit $m/R<<1$ we recover the classical expression $W=\omega R$.
The geodesics are null, timelike or spacelike if $W=1, <1, >1$ respectively.

Let us now focus on the radial motion of test particles. 
First of all, notice that as a consequence of the symmetry (spherical and time--independence) we have three constants of motion  which are energy and angular momentum (magnitude and direction), defined respectively by, (on the plane $\theta=\pi/2$),

\begin{equation}
\frac{\partial {\cal L}}{\partial \dot t}=constant\equiv E=-\dot t\left(1-\frac{2m}{R}\right),
\label{im1}
\end{equation}
\begin{equation}
\frac{\partial {\cal L}}{\partial \dot \phi}=constant\equiv L=\dot \phi R^2,
\label{im2}
\end{equation}
\begin{equation}
\frac{\partial {\cal L}}{\partial \dot \theta}=constant \equiv P_\theta=\dot \theta R^2.
\label{im3}
\end{equation}
Then  the first integral of (\ref{gs2})  may be written as 
\begin{equation}
\dot R^2= E^2-V^2,
\label{im3}
\end{equation}

 with
 \begin{equation}
V^2=\left(1-\frac{2}{y}\right)\left(\frac{\tilde L^2 }{y^2}+1\right)
\label{im4}
\end{equation}
where $y\equiv R/m$, $\quad$$\tilde L^2\equiv \frac{L^2}{m^2}$.

The above equation is the same equation (10) in \cite{1}. However in this reference it was used to study the motion inside the horizon, which obviously is incorrect (the potential $V$ given by (\ref{im4}) is correct but valid only outside the horizon).

For the motion along the axis $\theta=0$ we have  $L=0$, then for the value of energy given in Figure 1, all possible radial geodesics (for $R>2m$) extend  between the horizon (the vertical line)  and the value of $y$  where the horizontal line crosses the curve $V^2$ as given by  (\ref{im4}). We shall discuss about the behaviour of the particle for $R<2m$ in the next subsection. For larger values of $E$, such that $E>V$, unbounded trajectories are allowed.

 For $\theta=\pi/2$ and the values of energy $E$ and the angular momentum $\tilde L$ given in Figure 2,  the horizontal line crosses the curve $V^2$ as given by (\ref{im4}) (for $R>2m$) at two points, say $y_1, y_2$ ($y_2>y_1$).  Thus there are radial geodesics, outside the horizon,  in the interval  $y_1>y>2$, and    unbounded trajectories for $y>y_2$. The unstable circular geodesic corresponds to the value of $E=E_c$. The region inner to the horizon shall be considered in the next subsection.

All the  results above are well known, and apply for $R>2m$. 
 
\subsection{Geodesics for the hyperbolically symmetric black hole ($R<2m$)}
Let us now analyze  the geodesic structure for the region within the horizon, where we assume the space--time to be described by the hyperbolically symmetric solution (\ref{w3}).

Using (\ref{w3}) we obtain from (\ref{l2}) or (\ref{g11}):
\begin{equation}
\ddot t-\frac{2m\dot t \dot R}{R^2\left(\frac{2m}{R}-1\right)}=0,
\label{gsh1}
\end{equation}
\begin{eqnarray}
\ddot R&+&\frac{\dot R^2 m}{R^2(\frac{2m}{R}-1)}-\frac{m \dot t^2 (\frac{2m}{R}-1)}{R^2}\nonumber \\&-&R \dot \theta^2  \left(\frac{2m}{R}-1\right)-R\dot \phi ^2 \sinh^2 \theta  \left(\frac{2m}{R}-1\right)
=0,
\label{gsh2}
\end{eqnarray}
\begin{equation}
\ddot \theta R^2+2R\dot R \dot \theta-R^2\dot \phi ^2  \sinh\theta \cosh \theta=0,
\label{gsh3}
\end{equation}

\begin{equation}
\ddot \phi  +\frac{2\dot R \dot \phi}{R}+2\dot \phi \dot \theta \coth\theta=0.
\label{gsh4}
\end{equation}

Let us first consider circular geodesics along the $\phi$ direction. Thus $\dot R=\dot \theta=0$, and it follows from (\ref{gsh3})

\begin{equation}
R^2\dot \phi ^2  \sinh\theta \cosh \theta=0,
\label{gsh3b}
\end{equation}
from which we can see  that, unlike the case $R>2m$,  there are not circular geodesics in the $\phi$ direction, not even unstable ones.

Furthermore, from (\ref{gsh2}) it follows that
\begin{eqnarray}
\frac{m \dot t^2}{R^2}+R\dot \phi ^2 \sinh^2 \theta 
=0,\Rightarrow \omega^2=-\frac{m }{R^3 \sinh ^2\theta}
\label{gsh2b}
\end{eqnarray}
which is clearly unacceptable, and confirms the conclusion above.

Let us now consider geodesics along the $\theta$ direction, i.e. $ \dot R=\dot \phi=0$, then it follows from (\ref{gsh2}) 

\begin{eqnarray}
\frac{m \dot t^2}{R^2}+R \dot \theta^2 
=0,\Rightarrow \frac{\dot \theta^2}{\dot t^2}=-\frac{m}{R^3},
\label{gsh2c}
\end{eqnarray}
implying that there are not geodesics exclusively along the $\theta$ direction.

More generally if we assume $\dot R=0$ then it follows from  (\ref{gsh2}) 
\begin{eqnarray}
\frac{m \dot t^2}{R^2} +R \dot \theta^2  +R\dot \phi ^2 \sinh^2 \theta 
=0,
\label{gsh2a}
\end{eqnarray}
implying that no motion is possible unless $\dot R\neq 0$.

If we assume that $\dot \phi=0$ then it follows at once from (\ref{gsh3}) that
\begin{equation}
\dot \theta R^2=constant\Rightarrow P_\theta=constant,
\label{constheta}
\end{equation}

whereas, if we assume $\dot \theta=0$, then it follows from (\ref{gsh3}) that $\dot \phi=0$.

Next, let us assume that at some given initial $s=s_0$ we have $\dot \theta=0$, then it follows at once from (\ref{gsh3}) that such a condition will propagate in time only if $\theta=0$ or $\dot \phi=0$. In other words, any $\theta=constant$ trajectory is unstable except $\theta=0$, unless $\dot \phi=0$. It is worth stressing the difference between this case and the situation in the $R>2m$ region (see (\ref{gs3})).

Thus only the following cases are allowed:
\begin{enumerate}
\item Purely radial geodesics $\dot R\neq 0$, $\dot \theta=\dot\phi=0$.
\item Geodesics in the $R, \theta$ plane (i.e. $\dot \phi=0, \dot R, \dot \theta \neq 0$)
\item The general case $\dot R, \dot \theta,\dot\phi\neq 0$.
\end{enumerate}

\begin{figure}[h]
\includegraphics[scale=0.7]{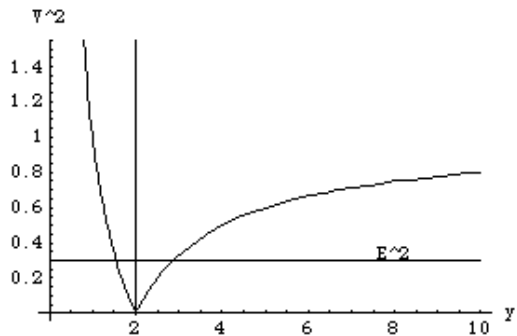}
\caption {\it  $V^2$ as function of $y$ for $\theta=0$. The vertical line is the horizon. The horizontal line corresponds to the value of $E^2=.315$}
\end{figure}

\begin{figure}[h]
\includegraphics[scale=0.7]{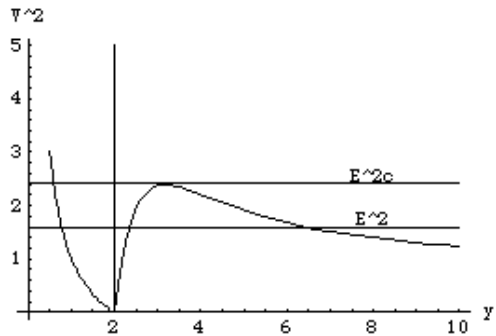}
\caption {\it  $V^2$ as function of $y$ for $\theta=\pi/2$, and $\tilde L^2=55$. The vertical line is the horizon. The lower horizontal line corresponds to the value of $E^2=1.71$. The higher horizontal line corresponds to the value of $E^2_c=2.42$}
\end{figure}
 
 Let us first consider  the radial motion of test particles inside the horizon.
As, for the region exterior to the horizon, we have two constants of motions which are energy and angular momentum, defined respectively by

\begin{equation}
\frac{\partial {\cal L}}{\partial \dot t}=constant\equiv E=\dot t\left(\frac{2m}{R}-1\right),
\label{im1}
\end{equation}
\begin{equation}
\frac{\partial {\cal L}}{\partial \dot \phi}=constant\equiv L=-\dot \phi R^2 \sinh^2\theta,
\label{im2}
\end{equation}

however,  the canonical momentum $P_\theta$ now  is not conserved, unless $\dot \phi=0$, 
\begin{equation}
\frac{\partial {\cal L}}{\partial \dot \theta}\equiv P_\theta=-\dot \theta R^2.
\label{im3}
\end{equation}

For the radial motion along the  symmetry  axis $\theta=0$, both $L$ and $P_\theta$ vanish. Also, as mentioned before, if we assume that at some initial time $\dot \theta=0$, then the trajectory along the $\theta =0$ axis will be stable in time.  

Then, the first integral of (\ref{gsh2}) with $\dot \phi=\dot \theta=0$ reads ,

\begin{equation}
\dot R^2= E^2-V^2,
\label{im3n}
\end{equation}

  with
 \begin{equation}
V^2=\left(\frac{2}{y}-1\right),
\label{im4n}
\end{equation}
where $y\equiv R/m$. 

The above equation is the same equation (15) in \cite{1}. However in this latter reference it was used to study the motion outside the horizon, which obviously is incorrect. Again, the potential $V$ given by (\ref{im4n} is correct but valid only  inside the horizon.

As we see from Figure 1,  for the given value of $E$, the test particle inside the horizon never reaches the center, moving between the point where the horizontal line crosses $V^2$ (as given by (\ref{im4n})) and the horizon. In principle the particle may cross the horizon and bounces back at the  point (outside the horizon) where the horizontal line crosse $V^2$ (as given by (\ref{im4}).

Thus for this particular value of energy we have a bounded trajectory with extreme points at both sides of the horizon. For sufficiently large (but finite)  values of energy, the particle moves between a point close to, but at finite distance from the center and $R\rightarrow \infty$.

Two main conclusions emerge at this point, for the test particle moving under the conditions stated above:
\begin{enumerate}
\item The particle never reaches the center, approaching it asymptotically as $E \rightarrow \infty$. 
\item The particle may cross the horizon, not only inwardly but also outwardly.
\end{enumerate}
The sharp difference between this behaviour of the test particle inside the horizon, and the corresponding behaviour in the ``classical'' picture of the Schwarzschild black hole, does not need to be further emphasized.

Let us now consider the radial motion on the  $\theta=\pi/2$ plane. First of all we observe that, as mentioned before, if we want to remain on that plane,  inside the horizon, we must have $\dot \phi=L=0$, and the situation is very similar to the case $\theta=0$, with one important difference; now the trajectory is unstable against perturbations of the angular momentum, and we should expect the particle to leave the $\theta=\pi/2$ plane. Nevertheless, for sake of completeness we have also plotted this case in figure 2.

Three main differences between this case and the situation for $R>2m$, deserve to be emphasized.
\begin{enumerate}
\item For $R>2m$ the motion on the plane $\theta=constant\neq 0$  is stable.
\item for $R<2m$ the motion on the plane $\theta=constant\neq 0$ requires $L=0$.
\item  Even if $L=0$, for $R<2m$, the trajectory will be unstable against perturbations of the angular momentum.
\end{enumerate}

These conclusions are qualitatively the same for any $\theta=constant \neq0$.

The results exhibited above show  that the motion along the $\theta=0$ axis is sharply different from other trajectories. More specifically, the instabilities of the motion for any $\theta=constant\neq0$  trajectory, implies that, unless the particle moves along $\theta=0$, all trajectories must  involve variations of $\theta$. Therefore, we shall next  find  the trajectories of test particles  on the plane $R-\theta$ for any $\phi=constant$, in which case the momentum $P_\theta$ is constant and $L=0$. It is worth noticing that due to  axial symmetry, the motion on any two dimensional slice $\phi=constant$ is invariant with respect to rotations around the symmetry axis. Therefore the restricted case $L=0$ provides the most  relevant physical information about the motion of the particle without integrating the full system of geodesic equations.

Then, the first integral of (\ref{gsh2}) becomes
\begin{equation}
\dot{R}=\sqrt{E^2-\left(\frac{2m}{R}-1\right)\left(\frac{P^2_\theta}{R^2}+1\right)}.
\label{geo1n}
\end{equation}

Since we are interested in the spatial trajectories, we use  
$$\dot{R}=-\frac{P_\theta}{R^2}R^\prime,\qquad R^\prime=\frac{dR}{d\theta},$$

to write (\ref{geo1n}) as 
\begin{equation}
  z^\prime=\frac{1}{k}\sqrt{E^2-(z-1)(k^2 z^2+1)}
\label{geo2n}
\end{equation}
where.
 $z=\frac{2m}{R}$,  and  $k=\frac{P_\theta}{2m}$, thus $z$ changes in the domain $(-\infty, 1]$.
We have integrated the equation above for a  wide range of values of the parameters $k,E$. The integration was carried out with the boundary condition that all trajectories coincide at $\theta =0$, $z=1$.

Two main results emerge  from all these models. On the one hand we found that  the test particle never crosses the horizon outwardly, approaching it as $k$ tends to zero, as expected from Figures 1,2. To illustrate this point we show  the  results of the numerical integration of (\ref{geo2n})  for the values indicated in the legends of Figures 3, 4, however this conclusion  holds  for  all possible trajectories with finite values of $E$ and $k$.  In these figures the axes cross at the origin (the center of symmetry)  and  the length of a line segment  from the center to any point on the curve is given by $z$ and the angle of this line with the horizontal axis is $\theta$. 
On the other hand, we found that the test particle never reaches the center, approaching it asymptotically as $k, E\rightarrow \infty$.

\begin{figure}[h]
\includegraphics[scale=0.7]{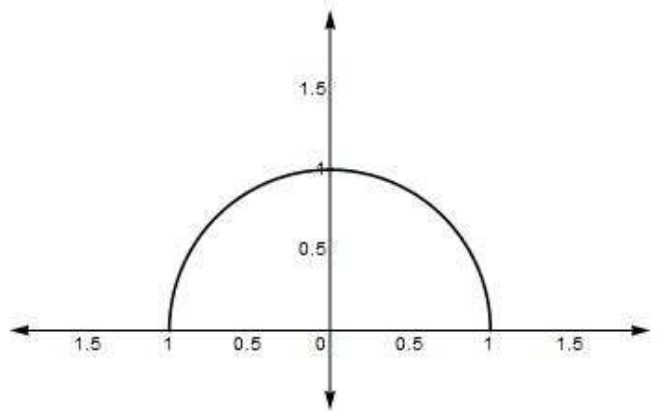}
\caption {\it  The trajectory of the test particle in the sub--space $R-\theta$, for $k=\frac{1}{10000}$ and $E=3$.}
\end{figure}

\begin{figure}[h]
\includegraphics[scale=0.7]{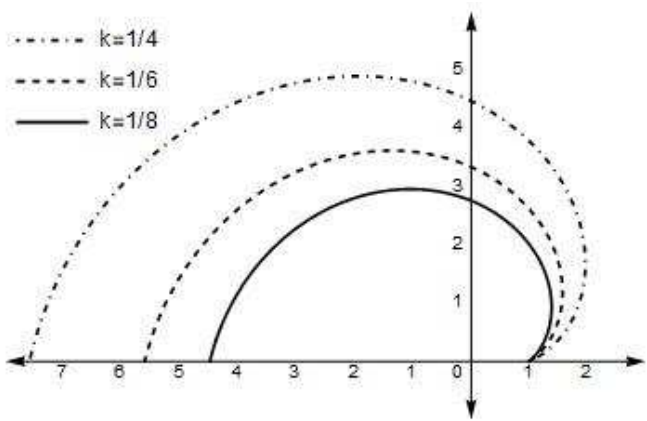}
\caption {\it  The trajectory of the test particle in the sub--space $R-\theta$ for three values of $k$, ($1/4, 1/6, 1/8$) and $E=3$.}
\end{figure}

In order to understand the results above, it is convenient to calculate  the four--acceleration of  a static observer in the frame of (\ref{w3}). We recall that a static observer is one whose four velocity $U^\mu$ is proportional to the Killing time--like vector \cite{Caroll}, i.e.
\begin{equation}
U^\mu=\left(\frac{1}{\sqrt{\frac{2m}{R}-1}}, 0, 0, 0 \right).
\label{1a}
\end{equation}
Then for the four--acceleration $a^\mu\equiv U^\beta U^\mu_{;\beta}$ we obtain for the region inner to the horizon
\begin{equation}
a^\mu=\left(0, -\frac{m}{R^2}, 0, 0 \right),
\label{2a}
\end{equation}
 whereas for the outer region described by (\ref{w2}) we obtain
 \begin{equation}
a^\mu=\left(0, \frac{m}{R^2}, 0, 0\right).
\label{3a}
\end{equation}

The physical meaning of (\ref{3a}) is clear, it  represents  the inertial radial acceleration outwardly  directed, which is necessary in order to maintain static the frame, by canceling the gravitational acceleration exerted on the frame. Since the former is directed radially outwardly, it means that the gravitational force is attractive, as expected.
 However, inside the horizon the four--acceleration as defined by (\ref{2a}) is directed inwardly, implying that a repulsive force is acting on the particle in that region. This remarkable fact explains the peculiarities of the orbits inside the horizon.
 
 The above discussion may be presented in an alternative format (see \cite{H} for details).
 Let us introduce a locally defined coordinate system  ($T, X, \theta, \phi)$ associated with a locally Minkowskian observer or, equivalently, a tetrad field associated with this Minkowskian observer, i.e.
 \begin{equation}
dX=\sqrt{-g_{RR}}dR \qquad dT=\sqrt{g_{tt}}dt
\label{a3}
\end{equation}
 then for a particle  instantaneously at rest inside the horizon, we have:
\begin{equation}
\frac{d^2X}{dT^2}=\frac{m}{R^2 \sqrt{(\frac{2m}{R}-1)}},
\label{a4}
\end{equation}
where (\ref{w3}) and (\ref{gsh2}) have been used.
 From the above equation, the repulsion experimented by the particle is clearly established.
 
 For the Schwarzschild solution (\ref{w2}) for $R >2 m$, the corresponding expression reads:
 \begin{equation}
\frac{d^2X}{dT^2}=-\frac{m}{R^2 \sqrt{(1-\frac{2m}{R})}},
\label{a5}
\end{equation}
indicating the attractive nature of the gravitation force in that region.
 \section{Gyroscope precession of a gyroscope along a circular, non--geodesic  path }
 We shall now calculate the precession of a gyroscope moving along a circular trajectory inside the horizon. Since, as we have already mentioned, no circular geodesics exist in that region, the trajectory of the gyroscope cannot be a geodesic. This calculation can be performed in different ways, here we shall use the Rindler-Perlick method \cite{Rindler}, which we find particularly suitable for our purpose.

This method consists in transforming the angular coordinate $\phi$ by
\begin{equation}
\phi=\phi^{\prime}+\omega t,
\end{equation}
where $\omega$ is a constant. Then the original frame is replaced by a
rotating frame. The transformed metric is written in a canonical form,
\begin{equation}
ds^2=-e^{2\Psi}(dt-\omega_idx^i)^2+h_{ij}dx^idx^j,
\end{equation}
with latin indexes running from 1 to 3 and $\Psi, \omega_i$ and $h_{ij}$
depending on the spatial coordinate $x^i$ only (we are omitting primes). Then,
it may be shown that the rotation three
vector $\Omega^i$ of the congruence of world lines $x^i=$constant is given
by \cite{Rindler},
\begin{eqnarray}
\Omega^i=\frac{1}{2}e^\Psi(\det h_{mn})^{-1/2}\epsilon^{ijk}\omega_{k,j},
\end{eqnarray}
where the comma denotes partial derivative, and the  three vector ${\bold \Omega}$ is related to the  vorticity tensor $\omega_{\alpha \beta}$ by

\begin{equation}
\omega_{k j}=\Omega^l\eta_{lkj},
\end{equation}
where $\eta_{lkj}$=$(det h_{mn})^{1/2}\epsilon_{ikj}$ is the Levi--Civita tensor associated to the spatial metric $h_{mn}$.

It is clear from the above that, since $\Omega^i$ describes the rate of
rotation with respect to the proper time at any point at rest in the
rotating frame, relative to the local compass of inertia, then $-\Omega^i$
describes the rotation of the compass of inertia (the {\it gyroscope}) with
respect to the rotating frame. 

Thus let us consider a gyroscope moving  around the center along a circular orbit (non--geodesic), and let us calculate its  precession. Since $-{\bold \Omega}$ describes the precession of the gyroscope relative to the lattice, then after one revolution the orientation of the gyroscope, in the rotating frame, changes by

\begin{equation}
\Delta \phi^\prime =-\Omega e^\Psi \Delta t.
\label{gn}
\end{equation}
Obviously, the precession per revolution relative to the original system is:
\begin{equation}
\Delta \phi =\Delta \phi^\prime +2\pi.
\label{gn3}
\end{equation}
The case of the Schwarzschild metric ($R>2m$) has been calculated in \cite{Rindler} for the $\theta=\pi/2$ plane. They obtain for the magnitude of the vorticity ${\bold \Omega}$ (notice that a $\omega$ is missing in the equation (38) in \cite{Rindler}):
\begin{equation}
\Omega=\frac{\omega \left(1-\frac{3m}{R}\right)}{1-\frac{2m}{R}-R^2\omega^2},
\label{g1}
\end{equation}
and for the total precession $\Delta \phi$
\begin{equation}
\Delta \phi=-2\pi\left[\frac{\left(1-\frac{3m}{R}\right)}{\sqrt{1-\frac{2m}{R}-R^2\omega^2}}-1\right].
\label{g2}
\end{equation}

From the expressions above, we see that at $R=3m$ the orientation of the gyroscope is locked to the lattice ($\Omega=0$). In the region between the horizon and $R=3m$, if $\omega$ is sufficiently small so that the orbits are time--like, $\Omega$ becomes negative and the precession of the gyroscope is forward  even in the rotating frame ($\Delta \phi^\prime>0$). Thus the total precession $\Delta \phi$ exceeds $2\pi$.

However for  the region interior to the horizon, as described by (\ref{w3}) the situation is completely different. Indeed, retracing the same steps leading to (\ref{g1}) and (\ref{g2}) we obtain (for the plane $\theta=\pi/2$)
\begin{equation}
\Omega=\frac{ \omega \alpha \left(\frac{3m}{R}-1\right)}{\frac{2m}{R}-1-R^2\omega^2 \alpha^2},
\label{g3}
\end{equation}

\begin{equation}
\Delta \phi^\prime =-\frac{2 \pi \alpha \left(\frac{3m}{R}-1\right)}{\sqrt{\frac{2m}{R}-1-R^2\omega^2 \alpha^2}},
\label{g4}
\end{equation}

\begin{equation}
\Delta \phi =-2\pi\left[\frac{\alpha \left(\frac{3m}{R}-1\right)}{\sqrt{\frac{2m}{R}-1-R^2\omega^2 \alpha^2}}-1\right],
\label{g5}
\end{equation}
where $\alpha\equiv \sinh ({\pi/2})$.

As it is apparent from (\ref{g4}),  for sufficiently small $\omega$, so that the orbits are time--like, in the region inner to the horizon described by the metric (\ref{w3}), the precession of the gyroscope is retrograde in the rotating frame.  Obviously the total precession in the original frame is now smaller than $2\pi$, as it happens for the Thomas precession in Minkowski space--time (see eqs. (32, 33) in \cite{Rindler}).
\section{Conclusions}
In the classical picture of the Schwarzschild black hole, any particle inside the horizon is bound  to reach the center in a finite proper time interval. This  is the basic fact behind the ``classical'' black hole paradigm. However, as we have seen here, if we adopt the point of view proposed in \cite{1}, we find that the kinematic and dynamic properties of a test particle inside the horizon, are quite different. Indeed, not only are the test particles not condemned to  displace to the center, but they cannot reach the center for any finite value of energy as shown in figures (1, 2, 4). This fact is brought about by the existence of a repulsive force within the horizon, that pushes  the test particle away from the center. 

Besides the feature commented above, there is another  important difference with respect to the ``classical'' picture. It consists in the fact that the particles inside the horizon may in principle leave that region along the axis $\theta=0$. Thus, the particle may come from $R\rightarrow \infty$ crosses the horizon, bounces back before reaching the center and crosses the horizon outward. As we have seen this can be done only along this axis, all other trajectories, as illustrated by figures (2, 3), never cross the horizon. This point was already emphasized in \cite{1}, although it must be stressed that other conclusions, concerning the motion of test particles presented in \cite{1} are erroneous due to an incorrect use of equations (\ref{im3}) and (\ref{im3n}). Also, it is worth noticing that  it is possible that a quantum theory would permit a particle to tunnel across the horizon for $\theta\neq 0$.

Finally we have seen that the precession  of a gyroscope moving along  a circle  inside the horizon is retrograde, whereas, close to the horizon,  but at the outside of it ($R=2m+\epsilon$, where $\epsilon<<2m$ and positive), the precession is forward.

Before closing this section we would like to raise two questions, and to speculate about their possible answers.

\begin{enumerate}
\item What is the physical origin of the repulsion experienced by the test particle inside the horizon?
\item What could be the observational consequences of the fact that the test particle could leave the horizon along the $\theta=0$ axis.?
\end{enumerate}

With respect to the first question,  let us mention that repulsive forces in the context of general relativity have been reported  before, in many different scenarios, (see \cite{repulsive3, papapetrou, repulsive2, repulsive4, re1, repulsive5, H, maluf, s1}). However, neither of these references provides a satisfactory physical explanation about the origin of such an effect. Although this requires a deep and careful analysis, which is beyond the scope of this manuscript, we speculate that  the repulsion might be related to quantum vacuum of the gravitational field. 
 
 With respect to the second question, we speculate that the hyperbolically symmetric black hole might be invoked to explain extragalactic relativistic jets.  
 
 Indeed, relativistic jets are highly energetic phenomena which have been observed in many systems (see \cite{Margon, Sams} and references therein), usually associated with the presence of a compact object, and exhibiting a high degree of collimation. The spin and the magnetic field of a compact object  are some of the  many mechanisms proposed so far to explain this phenomenon \cite{Blandford}. However, no consensus has been reached until now, concerning the basic mechanism for its origin. Still worse, the basic physical ideas underlying the occurrence of these jets are hidden by  the great number of models available (see \cite{Blandford, B1, B2, M1, m2, s2} and references therein) and their complexity, implying a large number of assumptions. Under these circumstances, we speculate that  the possible ejection, and collimation,  of test particles along the $\theta=0$ axis, produced by the repulsive force acting within the horizon, could be considered as a possible engine behind the jets.
 
 To summarize: Even though  our proposal for the region inside the horizon  is highly ``heterodox'', the results ensuing from its adoption lead to specific observational   consequences and provide possible explanation to some so far unresolved astrophysical quandaries. It goes without saying that the final word about the physical viability of the hyperbolically symmetric black hole belongs to the observational evidence. Nevertheless, in the light of the results here presented we believe that until such evidence is provided, our proposal is significant enough as to consider it further. 
 \begin{acknowledgments}
This work was partially
supported by Ministerio de Ciencia, Innovacion y Universidades. Grant number: PGC2018--096038--B--I00, and  Junta de Castilla y Leon. 
 Grant number:  SA083P17.
\end{acknowledgments}
  
\end{document}